\documentclass{aa}

\usepackage{graphicx}

\begin{document}

\title{Upper limits to the SN1006 multi-TeV gamma-ray flux from H.E.S.S. observations}

\titlerunning{H.E.S.S. upper limits for SN1006}

\author{F. Aharonian\inst{1}
 \and A.G.~Akhperjanian \inst{2}
 \and K.-M.~Aye \inst{3}
 \and A.R.~Bazer-Bachi \inst{4}
 \and M.~Beilicke \inst{5}
 \and W.~Benbow \inst{1}
 \and D.~Berge \inst{1}
 \and P.~Berghaus \inst{6} \thanks{Universit\'e Libre de 
 Bruxelles, Facult\'e des Sciences, Campus de la Plaine, CP230, Boulevard
 du Triomphe, 1050 Bruxelles, Belgium}
 \and K.~Bernl\"ohr \inst{1,7}
 \and C.~Boisson \inst{8}
 \and O.~Bolz \inst{1}
 \and C.~Borgmeier \inst{7}
 \and F.~Breitling \inst{7}
 \and A.M.~Brown \inst{3}
 \and J.~Bussons Gordo \inst{9}
 \and P.M.~Chadwick \inst{3}
 \and L.-M.~Chounet \inst{10}
 \and R.~Cornils \inst{5}
 \and L.~Costamante \inst{1,20}
 \and B.~Degrange \inst{10}
 \and A.~Djannati-Ata\"i \inst{6}
 \and L.O'C.~Drury \inst{11}
 \and G.~Dubus \inst{10}
 \and T.~Ergin \inst{7}
 \and P.~Espigat \inst{6}
 \and F.~Feinstein \inst{9}
 \and P.~Fleury \inst{10}
 \and G.~Fontaine \inst{10}
 \and S.~Funk \inst{1}
 \and Y.A.~Gallant \inst{9}
 \and B.~Giebels \inst{10}
 \and S.~Gillessen \inst{1}
 \and P.~Goret \inst{12}
 \and C.~Hadjichristidis \inst{3}
 \and M.~Hauser \inst{13}
 \and G.~Heinzelmann \inst{5}
 \and G.~Henri \inst{14}
 \and G.~Hermann \inst{1}
 \and J.A.~Hinton \inst{1}
 \and W.~Hofmann \inst{1}
 \and M.~Holleran \inst{15}
 \and D.~Horns \inst{1}
 \and O.C.~de~Jager \inst{15}
 \and I.~Jung \inst{1,13} \thanks{now at Washington Univ., Department of Physics,
 1 Brookings Dr., CB 1105, St. Louis, MO 63130, USA}
 \and B.~Kh\'elifi \inst{1}
 \and Nu.~Komin \inst{7}
 \and A.~Konopelko \inst{1,7}
 \and I.J.~Latham \inst{3}
 \and R.~Le Gallou \inst{3}
 \and A.~Lemi\`ere \inst{6}
 \and M.~Lemoine \inst{10}
 \and N.~Leroy \inst{10}
 \and T.~Lohse \inst{7}
 \and A.~Marcowith \inst{4}
 \and C.~Masterson \inst{1,20}
 \and T.J.L.~McComb \inst{3}
 \and M.~de~Naurois \inst{16}
 \and S.J.~Nolan \inst{3}
 \and A.~Noutsos \inst{3}
 \and K.J.~Orford \inst{3}
 \and J.L.~Osborne \inst{3}
 \and M.~Ouchrif \inst{16,20}
 \and M.~Panter \inst{1}
 \and G.~Pelletier \inst{14}
 \and S.~Pita \inst{6}
 \and G.~P\"uhlhofer \inst{1,13}
 \and M.~Punch \inst{6}
 \and B.C.~Raubenheimer \inst{15}
 \and M.~Raue \inst{5}
 \and J.~Raux \inst{16}
 \and S.M.~Rayner \inst{3}
 \and I.~Redondo \inst{10,20}\thanks{now at Department of Physics and
Astronomy, Univ. of Sheffield, The Hicks Building,
Hounsfield Road, Sheffield S3 7RH, U.K.}
 \and A.~Reimer \inst{17}
 \and O.~Reimer \inst{17}
 \and J.~Ripken \inst{5}
 \and L.~Rob \inst{18}
 \and L.~Rolland \inst{16}
 \and G.P.~Rowell \inst{1}
 \and V.~Sahakian \inst{2}
 \and L.~Saug\'e \inst{14}
 \and S.~Schlenker \inst{7}
 \and R.~Schlickeiser \inst{17}
 \and C.~Schuster \inst{17}
 \and U.~Schwanke \inst{7}
 \and M.~Siewert \inst{17}
 \and H.~Sol \inst{8}
 \and R.~Steenkamp \inst{19}
 \and C.~Stegmann \inst{7}
 \and J.-P.~Tavernet \inst{16}
 \and C.G.~Th\'eoret \inst{6}
 \and M.~Tluczykont \inst{10,20}
 \and D.J.~van~der~Walt \inst{15}
 \and G.~Vasileiadis \inst{9}
 \and P.~Vincent \inst{16}
 \and B.~Visser \inst{15}
 \and H.J.~V\"olk \inst{1}
 \and S.J.~Wagner \inst{13}}

\institute{\scriptsize
Max-Planck-Institut f\"ur Kernphysik, P.O. Box 103980, D 69029
Heidelberg, Germany
\and
 Yerevan Physics Institute, 2 Alikhanian Brothers St., 375036 Yerevan,
Armenia
\and
University of Durham, Department of Physics, South Road, Durham DH1 3LE,
U.K.
\and
Centre d'Etude Spatiale des Rayonnements, CNRS/UPS, 9 av. du Colonel Roche, BP
4346, F-31029 Toulouse Cedex 4, France
\and
Universit\"at Hamburg, Institut f\"ur Experimentalphysik, Luruper Chaussee
149, D 22761 Hamburg, Germany
\and
Physique Corpusculaire et Cosmologie, IN2P3/CNRS, Coll{\`e}ge de France, 11 Place
Marcelin Berthelot, F-75231 Paris Cedex 05, France
\and
Institut f\"ur Physik, Humboldt-Universit\"at zu Berlin, Newtonstr. 15,
D 12489 Berlin, Germany
\and
LUTH, UMR 8102 du CNRS, Observatoire de Paris, Section de Meudon, F-92195 Meudon Cedex,
France
\and
Groupe d'Astroparticules de Montpellier, IN2P3/CNRS, Universit\'e Montpellier II, CC85,
Place Eug\`ene Bataillon, F-34095 Montpellier Cedex 5, France 
\and
Laboratoire Leprince-Ringuet, IN2P3/CNRS,
Ecole Polytechnique, F-91128 Palaiseau, France
\and
Dublin Institute for Advanced Studies, 5 Merrion Square, Dublin 2,
Ireland
\and
Service d'Astrophysique, DAPNIA/DSM/CEA, CE Saclay, F-91191
Gif-sur-Yvette, France
\and
Landessternwarte, K\"onigstuhl, D 69117 Heidelberg, Germany
\and
Laboratoire d'Astrophysique de Grenoble, INSU/CNRS, Universit\'e Joseph Fourier, BP
53, F-38041 Grenoble Cedex 9, France 
\and
Unit for Space Physics, North-West University, Potchefstroom 2520,
    South Africa
\and
Laboratoire de Physique Nucl\'eaire et de Hautes Energies, IN2P3/CNRS, Universit\'es
Paris VI \& VII, 4 Place Jussieu, F-75231 Paris Cedex 05, France
\and
Institut f\"ur Theoretische Physik, Lehrstuhl IV: Weltraum und
Astrophysik,
    Ruhr-Universit\"at Bochum, D 44780 Bochum, Germany
\and
Institute of Particle and Nuclear Physics, Charles University,
    V Holesovickach 2, 180 00 Prague 8, Czech Republic
\and
University of Namibia, Private Bag 13301, Windhoek, Namibia
\and
European Associated Laboratory for Gamma-Ray Astronomy, jointly
supported by CNRS and MPG
}

\authorrunning{Aharonian et al.}

\date{Received / Accepted}

\offprints{Conor.Masterson@mpi-hd.mpg.de \& \\ \hspace*{4cm}Gavin.Rowell@mpi-hd.mpg.de}

\newpage\abstract{Observations of the shell-type supernova remnant SN1006 have been
          carried out with the H.E.S.S.
          system of Cherenkov telescopes during 2003 (18.2h with two
          operating telescopes) and 2004 (6.3h with all four telescopes). 
          No evidence for TeV $\gamma$-ray emission
          from any compact or extended region associated with the remnant is seen and resulting upper limits
          at the 99.9\% confidence level are up to a factor 10 lower than previously-published
          fluxes from CANGAROO. 
          For SN1006 at its current epoch of evolution we give limits for a number of important global parameters.
          Upper limits on the $\gamma$-ray 
          luminosity (for $E$=0.26 to 10 TeV, distance $d$=2 kpc) of $L_\gamma < 1.7\times 10^{33}$ erg s$^{-1}$, 
          and the total energy 
          in corresponding accelerated protons, $W_p<1.6\times 10^{50}$ erg (for proton energies $E_p \sim 1.5$ to 60 TeV and 
          assuming the lowest value $n=0.05$ cm$^{-3}$ of the ambient target density discussed in literature) 
          are estimated. Extending this estimate
          to cover the range of proton energies observed in the cosmic ray spectrum up to the knee 
          (we take here $E_p \sim$ 1~GeV to 3~PeV, assuming a differential 
          particle index $-$2) gives $W_p<6.3\times 10^{50}$ erg.
          A lower limit on the post-shock 
          magnetic field of $B>25\mu$G results when considering the synchrotron/inverse-Compton framework 
          for the observed X-ray flux and $\gamma$-ray upper limits.

}

\maketitle

\section{Introduction}
Observations of shell-type supernova remnants (SNR) at multi-GeV to TeV $\gamma$-ray energies have long been motivated by the 
idea that they are the prime source of hadronic cosmic-ray (CR) acceleration in our galaxy
(see for example Drury, Aharonian \& V\"olk \cite{Drury:1} and Naito \& Takahara \cite{Naito:1}). The 
detection of non-thermal X-ray emission from the young shell-type SNR SN1006 
(Koyama et al. \cite{Koyama:1}, Allen et al. \cite{Allen:1}) 
suggests that SN1006 is a site of electron acceleration to multi-TeV energies. 
Subsequent arc-second resolution results from Chandra (Long et al. \cite{Long:1}, Bamba et al. \cite{Bamba:1}) 
reveal the presence of several bright non-thermal X-ray arcs concentrating in the NE and SW regions of the SNR, 
which likely trace the location of strong shocks where at least electrons are accelerated.
These, and similar X-ray results from a number of other SNR over the past ten years, have provided a priority list of such 
targets for ground-based $\gamma$-ray telescopes. The results of CANGAROO-I (Tanimori et al. \cite{Tanimori:1}), suggesting
TeV $\gamma$-ray emission from the NE rim of SN1006 supported the notion that this SNR is  
capable of electron acceleration up to energies $\sim100$ TeV, when interpreted
in the synchrotron/inverse-Compton (IC) framework. The TeV photon spectrum is compatible with a power law 
of photon index $\Gamma \sim$2.3 (for $dN/dE \sim E^{-\Gamma}$) in the energy range 1.5 to 20 TeV 
(Naito et al. \cite{Naito:2}, 
Tanimori et al. \cite{Tanimori:2}). Later data from the CANGAROO-II 10 metre telescope also revealed SN1006 as an
emitter of TeV $\gamma$-rays (Hara et al. \cite{Hara:1}), yielding a compatible energy spectrum. 
The stand-alone HEGRA~CT1 telescope has also reported a significant excess in
observations taken at very large zenith angles and therefore at a high energy threshold  $E>18$ TeV (Vitale et al. \cite{Vitale:1}).

The interpretation of these X-ray and ground-based $\gamma$-ray results generally involves the
electronic synchrotron/IC and/or hadronic $\pi^0$-decay channels. Either channel can dominate, depending on, for example, the
density of ambient matter $n$ and magnetic field strength $B$. 
Detailed sampling of the $\gamma$-ray spectra {\em and} morphology 
are required to disentangle the electronic and hadronic components. The H.E.S.S.  (High Energy Stereoscopic System) experiment, 
the first of the next-generation ground-based $\gamma$-ray detectors to utilise the stereoscopic technique, has 
observed SN1006 during 2003 and 2004. The prime motivation of these observations, 
apart from confirmation of the CANGAROO results, has been to determine
the nature of any 100 GeV to TeV $\gamma$-ray emission from SN1006. 

Operating in the Southern Hemisphere, the H.E.S.S. experiment consists of four 
identical Cherenkov telescopes each with mirror area $\sim$107 m$^2$
(Bernl\"ohr et al. \cite{Bern:1}, Cornils et al. \cite{Cornils:1}, Vincent et al. \cite{Vincent:1}). 
All four telescopes have been operating since December 2003. 
Here, we present results using H.E.S.S. data taken during 2003 (two telescopes operating in stereo but
with considerable dead time of $\sim$50\% due to the lack of a central trigger system \cite{Funk:1}).
The four-telescope mode of 2004 with the central trigger is the most sensitive configuration for H.E.S.S.
We update here previous results of Masterson et al. (\cite{Masterson:1}) which
utilised a part of the 2003 dataset analysed without stereoscopic reconstruction (i.e. mono-mode). 
The large field of view (FoV $\sim 5^\circ$ diameter) of H.E.S.S. easily encompasses the SN1006 shell, and the angular 
resolution of $\sim 0.1^\circ$ attained (event-by-event) permits a search for $\gamma$-ray sources from different regions 
associated with the SNR.

\section{Analysis \& Results}

H.E.S.S. observation runs of $\sim$28 min duration were taken using the so-called {\em wobble} mode in which the 
tracking position is 
displaced $\pm 0.5^\circ$ in declination with respect to the centre of SN1006 
(RA 15$^h$02$^m$48.4$^s$ Dec $-$41$^\circ$54$^\prime$42$^{\prime\prime}$ J2000.0). Runs were accepted for analysis if 
they met a number of quality control 
criteria based on the recorded CR rate, the number of malfunctioning pixels in each camera and also the tracking performance. 
In particular telescope tracking was accurate to $\sim$20 arcsec for these data and verified by comparing the locations
of bright stars in the camera fields of view. 
Table~\ref{tab:observations} summarises observations for those runs meeting quality criteria. 
Without correction for system deadtime the overall
observation times were 30h (2003, with $\sim 50$\% deadtime) and 7h (2004, with $\sim 10$\% deadtime).
A total of 18.2h (2003) and 6.3h (2004) livetime of ON source data (where SN1006 is in the FoV) were available for
analysis. 
\begin{table}
  \caption{Summary of H.E.S.S. observations of SN1006.}
  \begin{tabular}{ccccc}
   Obs.         & Runs & ON source    & $^a<z>$   & Telescopes \\ 
   Period       &      & livetime [h] &           &  in use    \\ \hline
   Mar 2003     & 18   & 4.1          & 23.9      &  2       \\
   Apr 2003     & 21   & 5.0          & 25.4      &  2       \\
   May 2003     & 25   & 9.1          & 23.3      &  2       \\
   \bf 2003 Total & \bf 65 & \bf 18.2  & \bf 24.0 &          \\ \hline
   May 2004     & 15   & 6.3          & 28.5      &  4       \\ 
   \bf All Total & \bf 80  & \bf 24.5  & \bf 25.0 &          \\  \hline
  \multicolumn{5}{l}{\tiny a: Average zenith angle [deg] of observations.}\\
  \end{tabular}
  \label{tab:observations}
\end{table}

Individual telescope events coincident in time were merged for stereo analysis, followed by 
Cherenkov image reduction. 
Image reduction employs image `cleaning' which removes camera pixels dominated by skynoise, flat fielding of the 
camera responses
using a LED flasher, and absolute conversion from pixel ADC counts to photoelectron units 
using conversion constants obtained from special low-illumination runs.
Image moments such as {\em width} and {\em length} according to Hillas (\cite{Hillas:1}) are used as a basis to reject the 
dominating CR background. Rejection of the CR background is achieved by application of image shape cuts
({\em mean-reduced-scaled-width} $MRSW$ and {\em mean-reduced-scaled-length} $MRSL$). A further cut, $\theta$, the 
difference in assumed and reconstructed event directions is also applied.
Event directions are reconstructed according to algorithm `1' as detailed by Hofmann et al. (\cite{Hofmann:1}).
For a point-like source the combination of cuts on shape, and direction 
$\theta_{\rm cut}<0.14^\circ$
allows for rejection of over 99.9\% of CRs whilst retaining $\sim$40\% of $\gamma$-rays above a threshold energy of 260~GeV
(2003) and ~110 GeV (2004).
The full data processing chain and CR rejection applied to these data were defined {\em a-priori} using Monte Carlo 
simulations of $\gamma$-ray and real CR events, and fully verified on the Crab Nebula. Details can be found in 
Aharonian et al. (\cite{Aharonian:10}).

\begin{figure*}
 \centering
 \includegraphics[width=14cm]{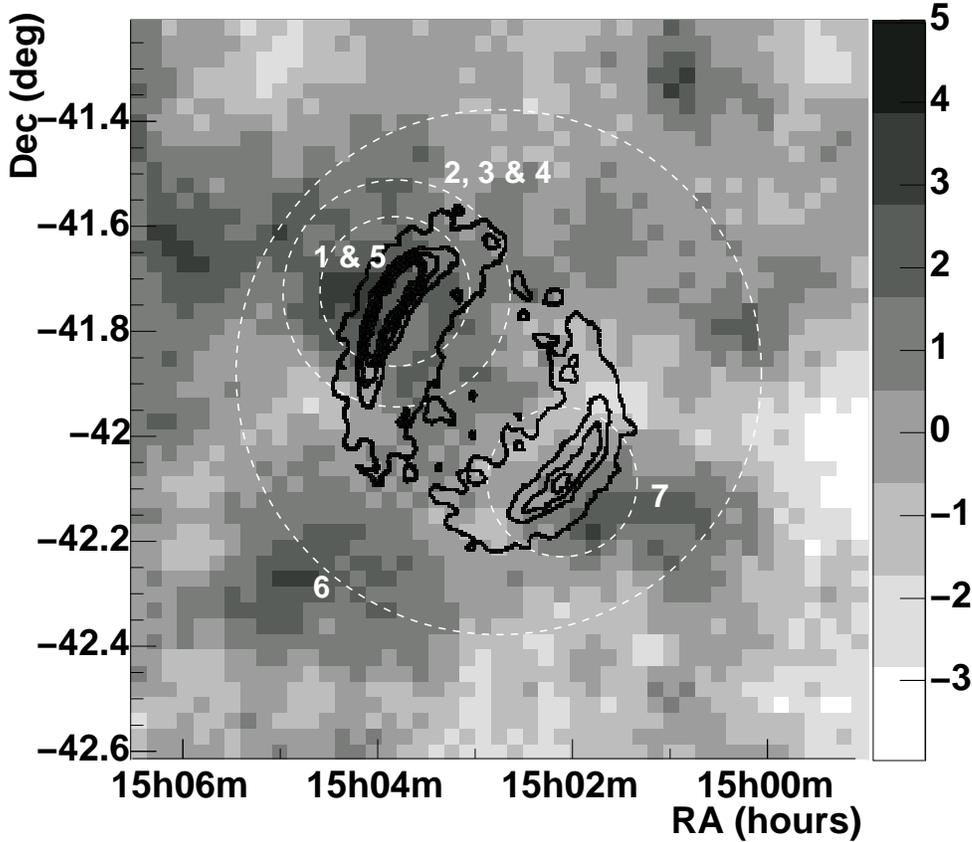}
 \caption{
   Skymap of excess significance (according to Eq. 17 of Li \& Ma
   (\cite{Li:1})) centred on SN1006.  Indicated are all {\em a-priori}
   integration regions used in these analyses (numbered according to
   the text and Table~\ref{tab:results}) and ASCA X-ray contours
   highlighting the X-ray emission from the NE and SW rims. The
   distribution of significances is fit by a Gaussian distribution with a
   mean $\mu = -0.016 \pm 0.004$ and sigma $\sigma = 1.035 \pm 0.003$. }
 \label{fig:skymap}
\end{figure*}

The skymap in Fig.~\ref{fig:skymap} presents the excess significance 
for $\gamma$-ray like events (i.e. after cuts) over the RA/Dec plane centred on SN1006. Data from 2003 and 2004 
were combined for these analyses.
At each bin position,
$\gamma$-ray-like events are summed within a circle of radius $\theta_{\rm cut}<0.14^\circ$ and the CR background is 
estimated from a ring region of radius 0.5$^\circ$ surrounding the source region. The ring region is chosen
to give a solid angle ratio $\alpha$ of 1:7 between the source and background regions respectively.
The skymap bins of Fig.~\ref{fig:skymap} are therefore correlated.
The cut  $\theta_{\rm cut}<0.14^\circ$ is appropriate for point-like $\gamma$-ray and marginally extended 
($\sim$ few arcmin) emission according to the H.E.S.S. point spread function (PSF).  
No significant excess suggesting $\gamma$-ray emission is evident in the skymap.
A very similar skymap is obtained when using the alternative {\em template} model (Rowell \cite{Rowell:1}) as 
a CR background estimate.
Event statistics and $\gamma$-ray flux upper limits at the 99.9\% confidence level, 
summarised in Table~\ref{tab:results}, 
have been calculated for a number of {\em a-priori} chosen locations based on the X-ray morphology as imaged 
by ASCA and the results of CANGAROO and HEGRA~CT1. The upper limits in the table have assumed a differential photon
index of $\Gamma$=2, and later (in Fig.~\ref{fig:compare_expt}) we give the upper limit band for a range of 
assumed indices (2 to 3). The {\em a-priori} locations, also indicated in Fig.~\ref{fig:skymap} are defined as:
\begin{enumerate}
 \item {\bf H.E.S.S. Point}:
 Upper limit on point-like emission ($E>0.26$ TeV) at the CANGAROO position using $\theta_{\rm cut}<0.14^\circ$ (appropriate 
 for the H.E.S.S. PSF). This is also used as an upper limit for the {\bf NE Rim} position.   
 \item {\bf CANGAROO Point}: Upper limit at the CANGAROO position using a $\theta_{\rm cut}<0.25^\circ$ cut
       appropriate for the CANGAROO 3.8m PSF and selecting events above their energy threshold ($E>$ 1.7 TeV).
 \item {\bf CANGAROO Point 2003}: As above using only 2003 data.
 \item {\bf CANGAROO Point 2004}: As above using only 2004 data.
 \item {\bf High Energy Point:} Excess events at the CANGAROO position using  $\theta_{\rm cut}<0.14^\circ$ 
      cut appropriate to the H.E.S.S. PSF but with an energy threshold $E>18$ TeV to compare with the HEGRA~CT1 result 
           (Vitale et al. \cite{Vitale:1}). Due to low statistics, the background is estimated from the 
          full FoV (radius 2.0$^\circ$) excluding
         the source region and corrected for differences in expected FoV efficiency. The dominant error comes from the on 
         count statistics.
 \item {\bf Whole SNR:} Entire SNR at the remnant centre using $\theta_{\rm cut}<0.50^\circ$, and H.E.S.S.
            energy threshold of ($E>$0.26 TeV).
 \item {\bf SW Rim:} As for position 1 but centered on the SE rim.
\end{enumerate}
\begin{table*}
 \centering
  \caption{Event statistics and flux upper limits (using the method of Feldman \& Cousins \cite{Feldman:1}, 
             assuming a differential photon index
           $\Gamma=2$) for a number of {\em a-priori}-chosen 
           regions associated with SN~1006:
           $s$ - source events; $b$ - CR background estimate; Normalisation factor $\alpha$; 
           See text for an explanation of the various regions. Except for limits 3 and 4, numbers are derived exclusively
           from combined 2003 and 2004 H.E.S.S. data.}
 \begin{tabular}{lllcccccc} 
  Region & $^{\rm a}$RA    &  $^{\rm a}$Dec    & $\theta_{\rm cut}$ & $s$  &  $b$  & $\alpha$ & $^{\rm b}$S & $^{\rm c} \phi^{99.9\%}$\\ \hline
  1. H.E.S.S. Point ($E>0.26$ TeV) & 15$^h$03$^m$48$^s$& $-$41$^\circ$45$^\prime$    & 0.14$^\circ$      & 1072 & 6854 & 0.150 &  1.19 & 2.19 \\
  \hspace{3mm}(also NE rim) &&& \\                                                                                    
  2. CANGAROO Point ($E>1.7$ TeV)        & 15$^h$03$^m$48$^s$& $-$41$^\circ$45$^\prime$    & 0.25$^\circ$      &   127&  792 & 0.153 &  0.48 & 0.34 \\
  3. CANG. Point ($E>1.7$ TeV)  & 15$^h$03$^m$48$^s$ & $-$41$^\circ$45$^\prime$    & 0.25$^\circ$      &   56 &  309 & 0.152 &  1.2  & 0.45 \\
  \hspace*{1cm}(2003 H.E.S.S. Data) \\
  4. CANG. Point ($E>1.7$ TeV)  & 15$^h$03$^m$48$^s$ & $-$41$^\circ$45$^\prime$    & 0.25$^\circ$      &   71 &  483 & 0.154 & -0.35 & 0.49 \\
 \hspace*{1cm}(2004 H.E.S.S. Data) \\
  5. H.E. Point ($E>18$ TeV) & 15$^h$03$^m$48$^s$ & $-$41$^\circ$45$^\prime$    & 0.25$^\circ$      &     0&   20 & 0.0069&    -  & 0.18$^{\rm d}$ \\ 
   6. Whole SNR ($E>0.26$ TeV)      & 15$^h$02$^m$48.4$^s$ & $-$41$^\circ$54$^\prime$42$^{\prime\prime}$ & 0.50$^\circ$      & 13358 & 63113 & 0.217 &  $-$2.8  & 2.39 \\ 
  7. SW Rim  ($E>0.26$ TeV)        & 15$^h$02$^m$4$^s$ & $-$42$^\circ$06$^\prime$3$^{\prime\prime}$  & 0.14$^\circ$      & 1042 & 6754 & 0.150 &  0.80 & 1.98 \\ \hline
  \multicolumn{9}{l}{\tiny a. RA \& Dec J2000.0 epoch}\\
  \multicolumn{9}{l}{\tiny b. Significance from Eq. 17 of Li \& Ma \cite{Li:1}, using normalisation factor $\alpha$}\\
  \multicolumn{9}{l}{\tiny c. $\phi^{99.9\%}_{ph} = $ 99.9\% integral upper limit ($\times 10^{-12}$ ph cm$^{-2}$s$^{-1}$)}\\
  \multicolumn{9}{l}{\tiny d. Dominant error is from $s$. We quote here the 99.9\% upper limit, which for a Poisson 
           of mean zero is 7 counts.}\\
  \end{tabular}
  \label{tab:results}
\end{table*}

\section{Discussion \& Conclusions}
\begin{figure}
 \centering
  \includegraphics[width=8.5cm]{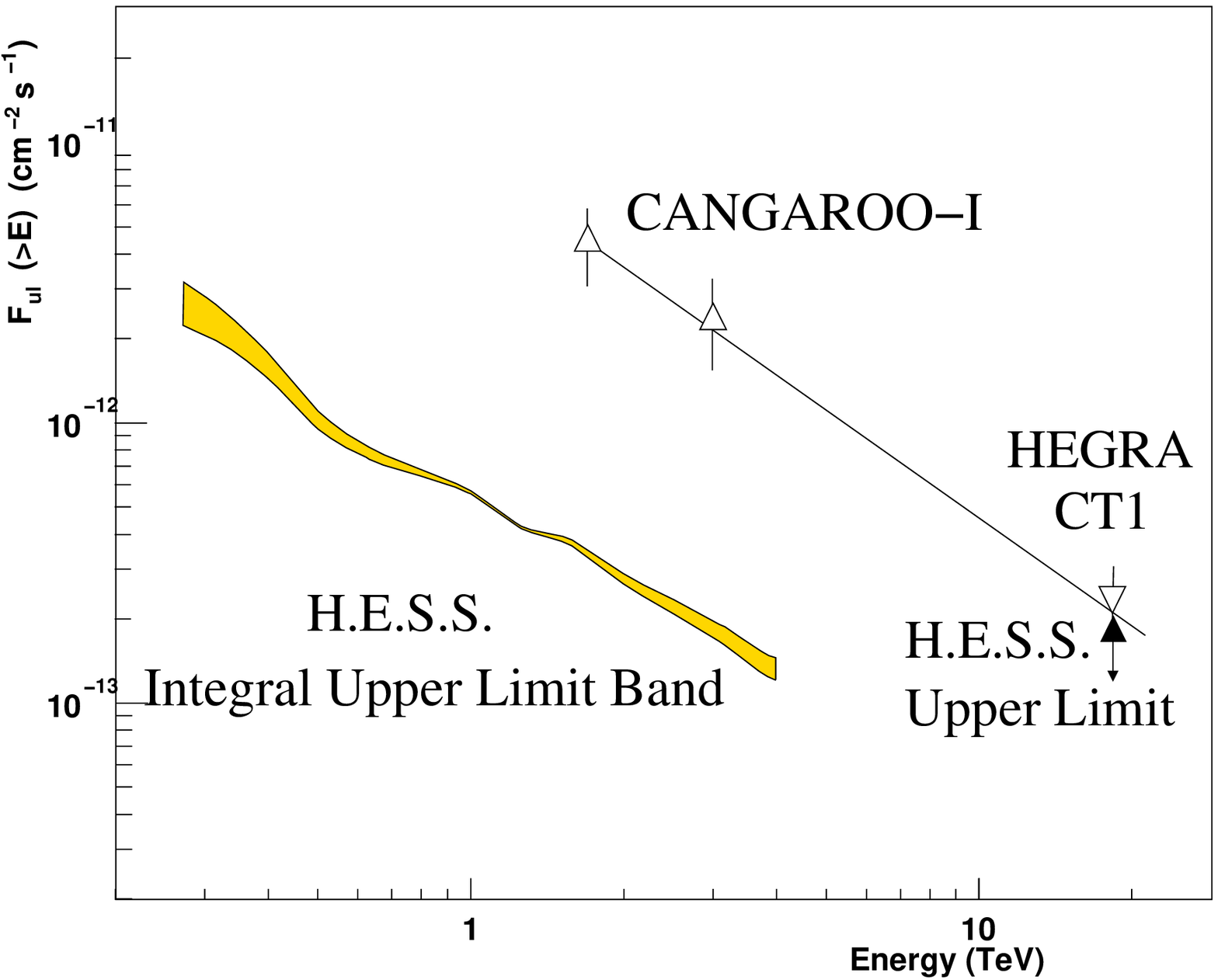}
  \includegraphics[width=8.5cm]{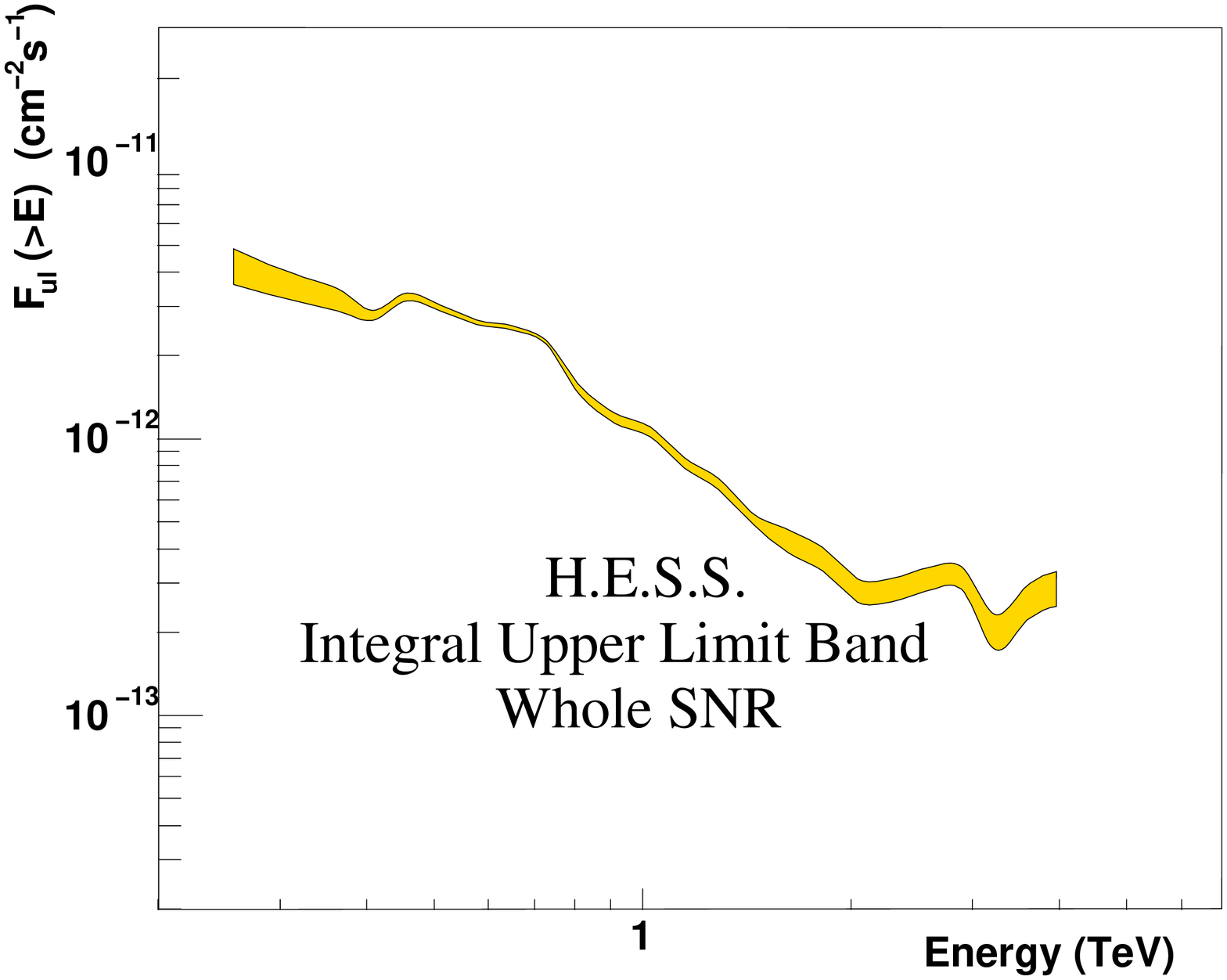}
 \caption{{\bf Top:} Comparison of experimental results for SN1006 from H.E.S.S. (integral upper limit 
      99.9\% c.l. band at the CANGAROO position using $\theta_{\rm cut}<0.23^\circ$, see Table~\ref{tab:results}),
      CANGAROO-I fluxes of Tanimori et al. (\cite{Tanimori:1}) 
      and the HEGRA CT1 flux of Vitale et al. (\cite{Vitale:1}).
      The solid line joins the two CANGAROO-I fluxes and is extended up to
      the HEGRA~CT1 flux. The high energy H.E.S.S. upper limit is shown for comparison with the HEGRA~CT1 flux. 
      The H.E.S.S. UL band arises when assuming a differential spectral index
      from $\Gamma$=2 to 3. {\bf Bottom:} H.E.S.S. upper limit band for the whole SNR, after cuts described in the text. The band
      arises when assuming a differential spectral index of $\Gamma$=2 to 3.}
 \label{fig:compare_expt}
\end{figure}

Comparing the H.E.S.S. upper limits with the fluxes from CANGAROO-I (1996 and 1997 observations) reveals a discrepancy
of a factor $\sim$10 (see Fig.~\ref{fig:compare_expt}). It is also clear that no $\gamma$-ray emission arises in 
just one of the years that H.E.S.S. has observed, 
2003 nor 2004 (see results given in Table~\ref{tab:results} for the CANGAROO position, 2003 and 2004 respectively).
Thus H.E.S.S. observations do not confirm the previously-published CANGAROO fluxes for SN1006. 
One would have to invoke a $\gamma$-ray flux variation over timescales less than 1 percent of the SNR age of 
    an order of magnitude, in order to explain the non-detection by H.E.S.S. This is however quite unlikely to occur 
    within the long-accepted diffusive shock acceleration framework for particle acceleration in SNR.

The HEGRA~CT1 telescope observed SN1006 at very large zenith angles
and obtained a preliminary flux of $F$ (E$>18$ TeV) = 2.5$\pm0.5_{\rm
  stat}\times$10$^{-13}$ ph cm$^{-2}$s$^{-1}$ (Vitale et al. \cite{Vitale:1}).  
We applied a matching high threshold cut to H.E.S.S.
data and obtained zero counts on source, (with $s-\alpha b$=$-$0.14 excess events).
The 99.9\% upper limit on this is a Poisson distribution with a mean of 7 events (we take here the
dominant error in the on source counts $s$ and quote the Poisson error).
In comparing with the HEGRA~CT1 result one must also take into consideration 
the systematic uncertainty on this flux, estimated to be $\geq$35\% (Vitale et al. \cite{Vitale:1}).
Given our exposure and considering both statistical and (uncorrelated) systematic uncertainties,
the HEGRA CT1 flux would yield 11$\pm$6 excess counts if seen by H.E.S.S. The
lower fluctuated value, at 5 counts, is consistent with our observed zero counts with a
chance probability of $\sim 7\times 10^{-3}$.  
We also include in Fig.~\ref{fig:compare_expt} the 99.9\% upper limit 
for the whole SNR as a function of energy. This allows comparison in 
cases where emission from models is predicted from regions not 
necessarily restricted to the NE and SW rims.

The H.E.S.S. upper limits permit estimation of important parameters for SN1006, in the framework
of this source as a particle accelerator. 
One can firstly estimate an upper limit to the $\gamma$-ray luminosity as $L_\gamma = 4\pi F d^2$ erg s$^{-1}$   
for a $\gamma$-ray flux $F$ and source distance $d$.
Distance estimates for SN1006 are in the range $d\sim$0.7 to 2.0 kpc,
 (see Allen et al. \cite{Allen:1} for a recent summary of measurements). Taking the H.E.S.S. upper limit for the whole 
SNR ($E>0.26$ TeV), distance $d=2.0$ kpc, assuming a differential photon spectrum of
$\Gamma=2$ and energy range 0.26 to 10 TeV gives 
$L_\gamma < 1.7 \times 10^{33}$ erg s$^{-1}$. 
This applies to the present epoch 
of SNR evolution for SN1006 which is likely in the early Sedov phase (Berezhko et al. \cite{Berezhko:1}).
We also estimate the energy in the corresponding accelerated protons $W_p$ 
over the SNR lifetime as $W_p \sim L_{\gamma} \tau_{pp}$  erg. The $W_p$ estimate applies to protons of energy 
$E_p \sim 1.5$ to 60 TeV, and accounts for the average inelasticity ($\sim 0.17$) of converting energy 
from protons to $\gamma$-rays in this scheme. The rather energy independent characteristic 
lifetime  $\tau_{pp}$ of accelerated protons interacting with an ambient matter of density $n$ cm$^{-3}$
($p+p \rightarrow \pi^0\,+\,X \rightarrow 2\gamma\, +\, X$) is $\tau_{pp} \sim 4.5\times10^{15}$ ($n$/cm$^{-3}$)$^{-1}$ sec. 
With $L_\gamma$ derived above we arrive at $W_p < $7.8$\times 10^{48}$
($n$/cm$^{-3}$)$^{-1}$ erg. Values for $n$ are in the range $n=0.05$ to $\sim 0.3$ cm$^{-3}$ based on X-ray and optical 
observations. Adopting the lowest value $n=0.05$ cm$^{-3}$ of the range of densities yields the 
upper limit $W_p<1.6\times 10^{50}$ erg.
The choice of $n$ in the $W_p$ estimate should reflect the average density of ambient matter inside the remnant undergoing
interaction with the downstream cosmic-rays (which are assumed to be confined inside the SNR).
A lower limit on this value would be that of the ambient matter. We can account for the fact that the energy band
of H.E.S.S. observations corresponding to protons ($\sim1.5$ to 60 TeV) is a fraction $\sim0.25$ of the 
total expected energy in protons over a much wider range of energies covering the cosmic-ray spectrum
up to the knee (we assume here $\sim$ 1 GeV to 3 PeV with a differential
particle index of $-$2). Given this, an upper limit on the full proton budget would be $W_p<6.3\times 10^{50}$ erg.   

Under the synchrotron/IC (scattering on the ubiquitous cosmic microwave background) scenario, knowledge of the 
synchrotron X-ray flux $f_x$ and TeV IC flux $f_\gamma$ arising from the {\em same} electrons permits a clear 
constraint on the $B$ field in 
the shock-compressed regions (downstream from the shock) of the SNR according to 
$f_\gamma(E_\gamma)/f_x(E_x) \sim 0.1(B/10\mu \rm G)^{-2} \xi$ 
(Aharonian et al. \cite{Aharonian:5}). It is assumed here that the X-ray and $\gamma$-ray emitting regions are of the 
same size, and thus the filling factor $\xi\sim1$. Such a case arises for high $B$ fields when electrons rapidly
cool at their synchrotron production regions. High values for 
$B>40\mu$G are in fact implied by X-ray observations and subsequent interpretation in the diffusive shock acceleration 
framework (Allen et al. \cite{Allen:1}, Berezhko et al. \cite{Berezhko:1,Berezhko:2}, Bamba et al. \cite{Bamba:1}, 
Yamazaki et al. \cite{Yamazaki:1}, Ksenofontov et al. \cite{Ksenofontov:1}).
With the H.E.S.S. upper limits on the $\gamma$-ray emission, one can therefore estimate a lower limit 
on the $B$-field. The synchrotron ($E_x$) and IC
($E_\gamma$) photon energies are coupled according to $E_\gamma \sim 1.5 (E_x /0.1 {\rm keV})(B/10 \mu \rm G)^{-1}$ TeV, 
requiring that the fluxes $f_x$ and $f_\gamma$ in appropriate energy ranges be compared. 
Comparing the available X-ray energy flux 
($E=0.1$ to 2 keV) $f_x = 1.42\times10^{-10}$ erg cm$^{-2}$s$^{-1}$  (Allen et al. \cite{Allen:1}) with that
of H.E.S.S. (upper limit energy flux for the whole SNR over an appropriate energy range E$\sim$1 to 10 TeV 
and assuming a $-$2.0 spectral index) at
$f_\gamma$=2.29$\times10^{-12}$ erg cm$^{-2}$s$^{-1}$ 
yields a lower limit of $B>25\mu$G. This lower limit on $B$ is consistent 
with values discussed earlier which result from comparisons of X-ray observations with detailed theory. 
In summary, H.E.S.S.
observations of SN1006 have not revealed evidence for TeV $\gamma$-ray emission. The 
resulting upper limits will be valuable in constraining further the parameters of $\gamma$-ray production in SN1006 
(see e.g. Aharonian \& Atoyan \cite{Aharonian:1}; Berezhko et al. \cite{Berezhko:1}, Ksenofontov et al. \cite{Ksenofontov:1}).

\begin{acknowledgements}
The support of the Namibian authorities and of the University of Namibia
in facilitating the construction and operation of H.E.S.S. is gratefully
acknowledged, as is the support by the German Ministry for Education and
Research (BMBF), the Max Planck Society, the French Ministry for Research,
the CNRS-IN2P3 and the Astroparticle Interdisciplinary Programme of the
CNRS, the U.K. Particle Physics and Astronomy Research Council (PPARC),
the IPNP of the Charles University, the South African Department of
Science and Technology and National Research Foundation, and by the
University of Namibia. We appreciate the excellent work of the technical
support staff in Berlin, Durham, Hamburg, Heidelberg, Palaiseau, Paris,
Saclay, and in Namibia in the construction and operation of the
equipment.

\end{acknowledgements}

\end{document}